\def\rfr#1{eq. (\ref{#1})}

\def\dert#1#2{\frac{{{d}}{#1}}{{{d}}{#2}}}

\def\virg#1{``#1''}

\def\eqi{\begin{equation}}
\def\eqf{\end{equation}}
\def\eqia{\begin{eqnarray}}
\def\eqfa{\end{eqnarray}}
\def\rp#1#2{{#1\over#2}} \def\lb#1{\label{#1}}

\def\bds#1{\boldsymbol{#1}}

\documentclass{ws-ijmpd}
\usepackage{latexsym,wasysym}
\usepackage{graphicx,epsfig}
\RequirePackage{color}

\begin{document}

\markboth{L. Iorio}
{Orbital effects of a time-dependent Pioneer-like acceleration}

%
%

\title{ORBITAL EFFECTS OF A TIME-DEPENDENT PIONEER-LIKE ANOMALOUS ACCELERATION}

\author{L. IORIO}

\address{Ministero dell'Istruzione, dell'Universit\`{a} e della Ricerca (M.I.U.R.)-Istruzione\\
International Institute for Theoretical Physics and
High Mathematics Einstein-Galilei.\\ Permanent address: Viale Unit\`{a} di Italia 68
Bari, (BA) 70125,
Italy.\\
e-mail: lorenzo.iorio@libero.it}

\maketitle

\begin{history}
\received{16 September 2011}
\end{history}

\begin{abstract}
We work out the impact that the recently determined time-dependent component of the Pioneer Anomaly (PA), if interpreted as an additional exotic acceleration of gravitational origin with respect to the well known PA-like constant one, may have on the orbital motions of some planets of the solar system. By assuming that it points towards the Sun, it turns out that  both the semi-major axis $a$ and the eccentricity $e$ of the orbit of a test particle would experience secular variations. For Saturn and Uranus, for which modern data records cover at least one full orbital revolution, such predicted anomalies  are up to $2-3$ orders of magnitude larger than the present-day accuracies in empirical determinations of their orbital parameters from the usual orbit determination procedures in which the PA was not modeled. Given the predicted huge sizes of such hypothetical signatures, it is unlikely that their absence from the presently available processed data can be attributable to an \virg{absorption} for them  in the estimated parameters caused by the fact that they were not explicitly modeled. The magnitude of a constant PA-type acceleration  at $9.5$ au cannot be larger than  $9\times 10^{-15}$ m s$^{-2}$ according to the latest observational results for the perihelion precession of Saturn.
\end{abstract}



\section{Introduction}
According to the latest analysis\cite{Tury011} of extended data records of the Pioneer $10/11$ spacecraft,
the small frequency drift\cite{And98,And02} (blue-shift) observed analyzing the navigational data of both the spacecraft, known as Pioneer Anomaly (PA),  may present a further time-dependent
component in addition to the well known  constant one. Both linear and  exponential models were proposed\cite{Tury011} for the PA; according to the authors of Ref.~\refcite{Tury011}, the exponential one is directly connected to  non-gravitational effects\cite{Toth09} since it takes into account the possible role of the on-board power generators suffering a radioactive decay.

In this letter we  work out the orbital effects of such a new term in the hypothesis that the time-dependent PA component is due to some sort of  long-range modification of the known laws of gravitation resulting in an additional anomalous acceleration with respect to the nearly sunward constant one, having magnitude\cite{And02}
\eqi \left|A_{\rm Pio}\right|=(8.74\pm 1.33)\times 10^{-10}\ {\rm m\ s^{-2}},\eqf in terms of which the constant part of the PA has often been interpreted. Indeed, in this case it should act on the major bodies of the solar system as well, especially those whose orbits lie in the  regions in which the PA manifested itself in its presently known form. In this respect, we will not consider the exponential model.
Recent studies\cite{Berto08,Riv09,Berto010,Riv010a,Riv010b,Fran011,Riv011}, partly preceding the one in Ref.~\refcite{Tury011}, pointed towards a mundane explanation of a large part of the PA in terms of non-gravitational effects pertaining the spacecraft themselves.
\section{The orbital effects of the linear model}
Since the anomalous acceleration is\cite{Tury011}
\eqi \dot A_{\rm Pio}\approx -2\times 10^{-11}\ {\rm m\ s^{-2}\ yr^{-1}},\eqf
the time-dependent linear component of the postulated PA-type acceleration\cite{Tury011} \eqi A= \left(t-t_0\right)\dot A_{\rm Pio}\lb{accela}\eqf
can be treated as  a small perturbation of the dominant Newtonian monopole $A_{\rm N}$ over timescales of the order of an orbital period $P_{\rm b}$ for all the planets of the solar system. Table \ref{pertu} explicitly shows this fact for Saturn, Uranus, Neptune and Pluto which move just in the spatial regions in which the PA perhaps started to appear (Saturn), or fully manifested itself (Uranus, Neptune, Pluto) in its  presently known form.
\begin{table}[ph]
\tbl{Comparison between the magnitudes of the Newtonian monopole accelerations $A_{\rm N}$ of the outer planets of the solar system and their putative Pioneer-type accelerations $A$ of \rfr{accela} over timescales comparable to their orbital periods $P_{\rm b}$. Also the values of the planetary semi-major axes $a$, in au, and eccentricities $e$ are displayed. The gravitational parameter of the Sun is $GM_{\odot}=1.32712\times 10^{20}$ m$^3$ s$^{-2}$.
\label{pertu}}
{\begin{tabular}{@{}cccccc@{}} \toprule
Planet & $a$ (au) & $e$ & $\left|A_{\rm N}\right|$ (m s$^{-2}$)  & $P_{\rm b}$  (yr) & $|A|$ (m s$^{-2}$) \\
\colrule
Saturn & $9.582$ & $0.0565$ & $6\times 10^{-5}$ & $29.457$ & $6\times 10^{-10}$  \\
Uranus & $19.201$ & $0.0457$ & $2\times 10^{-5}$ & $84.011$ & $2\times 10^{-9}$ \\
Neptune & $30.047$ & $0.0113$ & $6\times 10^{-6}$ &  $164.79$ & $3\times 10^{-9}$ \\
Pluto & $39.482$ & $0.2488$ & $4\times 10^{-6}$ & $247.68$ & $5\times 10^{-9}$ \\
\botrule
\end{tabular}}
\end{table}
Thus, the Gauss equations for the variation of the osculating Keplerian orbital elements\cite{Brou}, which are valid for any kind of disturbing acceleration $\bds A$, independently of its physical origin, can be safely used for working out the orbital effects of \rfr{accela}.
In particular, the Gauss equations for the semi-major axis $a$ and  eccentricity $e$ of the orbit of a test particle moving around a central body of mass $M$
are
\begin{equation}
\begin{array}{lll}
\dert a t & = & \rp{2}{n\sqrt{1-e^2}} \left[e A_R\sin f +A_{T}\left(\rp{p}{r}\right)\right],\\   \\
\dert e t  & = & \rp{\sqrt{1-e^2}}{na}\left\{A_R\sin f + A_{T}\left[\cos f + \rp{1}{e}\left(1 - \rp{r}{a}\right)\right]\right\}:
\end{array}\lb{Gauss}
\end{equation}
they allow one to work out the rates of changes of $a$ and $e$ averaged over one orbital period $P_{\rm b}$ as
\eqi\left\langle\dert\Psi t\right\rangle = \left(\rp{1}{P_{\rm b}}\right)\int_0^{P_{\rm b}}\left(\dert\Psi t\right)_{\rm K}dt,\ \Psi =a,e.\lb{media}\eqf In \rfr{media}
$\left(d\Psi/dt\right)_{\rm K}$ are the right-hand-sides of \rfr{Gauss} evaluated onto the unperturbed Keplerian ellipse.
In \rfr{Gauss} $A_R, A_T$ are the radial and transverse components of a the generic disturbing acceleration $\bds A$, $p\doteq a(1-e^2)$ is the semilatus rectum, $n\doteq \sqrt{GM/a^3}$ is the unperturbed Keplerian mean motion related to the orbital period by $n=2\pi/P_{\rm b}$, $G$ is the Newtonian constant of gravitation, and $f$ is the true anomaly.
Since the new data analysis\cite{Tury011} does not rule out the line joining the Sun and the spacecrafts as a direction for the PA, we will assume that \rfr{accela} is entirely radial, so that $A_R=A, A_T=0$.
 Using the eccentric anomaly $E$ as a fast variable of integration turns out to be computationally more convenient. To this aim, useful relations are
\eqi
\begin{array}{lll}
dt &=& \left(\rp{1-e\cos E}{n}\right)dE, \\ \\
t-t_0 &=& \left(\rp{E-e\sin E -E_0+ e \sin E_0}{n}\right), \\ \\
\sin f &=& \rp{\sqrt{1-e^2}\sin E}{1-e\cos E}.
%
\end{array}
\eqf
As a result, $a$ and $e$ experience non-vanishing secular variations
\eqi
\begin{array}{lll}
\left\langle \dert a t\right\rangle & = & -\rp{\dot A_{\rm P}a^3 e\left(2+e\right)}{GM}, \\ \\
\left\langle \dert e t\right\rangle & = & -\rp{\dot A_{\rm P}a^2 e\left(2+e\right)\left(1-e^2\right)}{2 GM}.
\end{array}\lb{rates}
\eqf
Notice that \rfr{rates} are exact in the sense that no approximations in $e$ were assumed. Moreover, they do not depend on $t_0$.

In order to make a meaningful comparison of \rfr{rates}
with the latest empirical results from planetary orbit determinations, we recall that modern data records cover at least one full orbital revolution for all the planets with the exception of Neptune and Pluto.
The author of Ref.~\refcite{Pit07},  in producing the EPM2006 ephemerides,  made a global fit of a complete suite of standard dynamical force models acting on  the solar system's major bodies to more than 400,000 observations of various kinds ranging over $\Delta t=93$ yr ($1913-2006$). Among the about $230$  estimated parameters, there are the planetary orbital elements as well.
According to Table 3 of Ref.~\refcite{Pit07}, the formal, statistical errors in $a$ for Saturn and Uranus are
\eqi
\begin{array}{lll}
\sigma_{a_{\saturn}}^{\rm (EPM2006)} &=& 4,256\ {\rm m}\\ \\
\sigma_{a_{\uranus}}^{\rm (EPM2006)} &=& 40,294\ {\rm m},\\ \\
\end{array}\lb{pita}
\eqf
so that
\eqi
\begin{array}{lll}
\sigma_{\dot a_{\saturn}}^{\rm (EPM2006)} &=& 46\ {\rm m\ yr^{-1}}\\ \\
\sigma_{\dot a_{\uranus}}^{\rm (EPM2006)} &=& 433\ {\rm m\ yr^{-1}},\\ \\
\end{array}\lb{ratesaer}
\eqf
can naively be inferred for their rates by simply dividing \rfr{pita} by $\Delta t$. The PA was not modeled in the EPM2006. It is important to remark that the figure for $\sigma_{a_{\saturn}}$ quoted in \rfr{pita} was obtained without processing the radiotechnical observations of the Cassini spacecraft.
According to \rfr{rates}, the putative PA-induced secular changes of the semi-major axes of Saturn and Uranus are
\eqi
\begin{array}{lll}
\left\langle\dot a_{\saturn}^{(\rm Pio)}\right\rangle &=& 42,505\ {\rm m\ yr^{-1}}\\ \\
\left\langle\dot a_{\uranus}^{(\rm Pio)}\right\rangle &=& 290,581\ {\rm m\ yr^{-1}}.\\ \\
\end{array}\lb{ratesa}
\eqf
These are about $3$ orders of magnitude larger than \rfr{ratesaer}: even by re-scaling the formal uncertainties of \rfr{ratesaer} by a factor of 10, the PA-type anomalous rates of \rfr{ratesa} would  still be about 2 orders of magnitude too large to have escaped from a detection.
Such conclusions are confirmed, and even enforced, by using the latest results published in Ref.~\refcite{Pit011} whose authors explicitly estimated secular changes of the semi-major axes of the first six planets with the EPM2010 ephemerides based on more than 635,000  observations  of different types over $\Delta t=97$ yr (1913-2010). The authors of Ref.~\refcite{Pit011}, whose goal was a different one, did not model the PA. They obtain\footnote{A $35.9\%$ correlation between $\dot a_{\saturn}$ and $a_{\saturn}$ was reported in Ref.~\refcite{Pit011}.}
\eqi \dot a^{\rm (EPM2010)}_{\saturn} = 13\ {\rm m\ yr^{-1}},\eqf which is 4 orders of magnitude smaller than the predicted value of \rfr{ratesa}.

Incidentally, we note that if \rfr{accela} acted on, say, the Earth, it would cause a variation in its semi-major axis as large as
\eqi \left\langle\dot a^{(\rm Pio)}_{\oplus}\right\rangle=14.5\ {\rm m\ yr^{-1}}\lb{prediterra}\eqf corresponding to a shift of \eqi\Delta a^{(\rm Pio)}_{\oplus}=1,348.5\ {\rm m}\lb{da}\eqf over $\Delta t=93$ yr. The formal, statistical uncertainty in the terrestrial semi-major axis is\cite{Pit07}
\eqi\sigma_{a_{\oplus}}^{\rm (EPM2006)}=0.138\ {\rm m},\lb{erra}\eqf i.e. about 4 orders of magnitude smaller than \rfr{da}. Even by re-scaling \rfr{erra} by a factor of 10, \rfr{da} would still be 3 orders of magnitude too large.
The EPM2010 ephemerides\footnote{A $0.6\%$ correlation between $\dot a_{\oplus}$ and $a_{\oplus}$ was reported in Ref.~\refcite{Pit011}.} yield\cite{Pit011}
\eqi \dot a_{\oplus}^{\rm (EPM2010)}=2\times 10^{-5}\ {\rm m\ yr^{-1}},\eqf i.e. 6 orders of magnitude smaller than the prediction of \rfr{prediterra}.
%

In principle, it may be argued that \rfr{accela} was not included in the mathematical models fitted to the planetary data in the orbit determination process, so that its signature may have  partly or totally been absorbed in the estimation  of, say, the planetary state vectors: thus, the entire observational record should  be re-processed by using ad-hoc  modified dynamical  force models explicitly including \rfr{accela} itself. However, this argument may have a validity especially when the magnitude of a putative anomalous effect of interest is  close to the accuracy of the orbit determination process: it is not the case here. Moreover, as far as the constant PA term is concerned, independent analyses\cite{Sta08,Fie09,Sta10} in which it was explicitly modeled as a radial extra-acceleration acting on the outer planets substantially confirmed its neat incompatibility with the observations, as suggested\footnote{Concerning the inner planets of the solar system, from an analysis of their determined orbital motions the authors of Ref.~\refcite{And02} pointed out that there was no room for a constant, sunward PA-like exotic acceleration acting on them.} in earlier  studies\cite{Iorio06,Iorio07}.

Concerning the possibility that the PA is present at Saturn, the authors of Ref.~\refcite{Tury011} yield an equivalent acceleration
\eqi A^{\left(\saturn\right)}_{\rm Pio}=(4.58\pm 11.80)\times 10^{-10}\ {\rm m\ s^{-2}}\lb{sat}\eqf
for its constant term. It is known that a radial constant extra-acceleration causes a secular precession of the perihelion\cite{Iorio06,San06,Ser06,Adk07}
given by \eqi \dot\varpi^{\rm (Pio)}=\rp{A_{\rm Pio}\sqrt{1-e^2}}{na}.\lb{pioprec}\eqf
The present-day accuracy in empirically constraining any anomalous precession $\Delta\dot\varpi$ of the perihelion of Saturn from the INPOP10a ephemerides, in which the PA was not modeled,
is\cite{Fie011}
\eqi\sigma_{\Delta\dot\varpi_{\saturn}}= 0.65\ {\rm milliarcsec\ cty^{-1}}.\lb{bounds}\eqf
Thus, \rfr{pioprec} and \rfr{bounds} yield
\eqi\left|A^{\left(\saturn\right)}_{\rm Pio}\right|\leq \rp{na}{\sqrt{1-e^2}}\sigma_{\Delta\dot\varpi_{\saturn}}=9\times 10^{-15}\ {\rm m\ s^{-2}},\eqf which is up to $5$ orders of magnitude smaller than \rfr{sat}.
\section{Conclusions}
If  the time-dependent linear part of the PA were a genuine effect of gravitational origin causing an extra-acceleration,  it would affect the orbital motions of the planets with secular variations of their semi-major axis and eccentricity if it were  oriented towards the Sun. The magnitude of such putative effects for Saturn and Uranus is up to $2-3$ orders of magnitude larger than the current accuracy in determining their orbital motions from the observations without modeling the PA itself. A partial removal of a PA-like signature may, in principle, have occurred in the standard parameter estimation procedure, so that the entire planetary data set should be re-processed with ad-hoc modified dynamical models explicitly including a hypothetical time-dependent PA-type acceleration as well.  However, supported by the results of analogous tests for the constant form of the PA actually implemented by independent teams of astronomers with different purposely modified dynamical models, we feel that it is unlikely that such a potential removal of a PA-type signature  can really justify the absence of such a huge effect in the currently available processed planetary data. Concerning the existence of a constant PA-type acceleration at heliocentric distances of about $9.5$ au as large as $1.6\times 10^{-9}$ m s$^{-2}$, latest results about the maximum allowed size of anomalies in the  perihelion precession of Saturn yield a figure 5 orders of magnitude smaller.



\end{document}